\documentclass[amscd,amssymb,onecolumn,showpacs,verbatim,11pt]{revtex4}
\def\dspace{\baselineskip=0.3 in}
\usepackage{graphicx}
\usepackage{dcolumn}
\usepackage{bm}

\input epsf
\begin{document}

\dspace

\title{\Large GENERALISED SECOND LAW OF THERMODYNAMICS FOR INTERACTING DARK ENERGY IN THE DGP BRANE WORLD}
\smallskip

\author{\bf~Jibitesh~Dutta $^{1,3}$\footnote{jdutta29@gmail.com,~jibitesh@nehu.ac.in}
and Subenoy~Chakraborty
$^2$\footnote{schakraborty@math.jdvu.ac.in} }

\smallskip

\affiliation{$^{1}$Department of Basic Sciences and Social
Sciences,~ North Eastern Hill University,~NEHU Campus, Shillong -
793022 ( INDIA )}

\affiliation{$^2$Department of Mathematics,~Jadavpur
University,~Kolkata-32, ( INDIA )}

\affiliation{$^3$Department of Mathematics,~ North Eastern Hill
University,~NEHU Campus, Shillong - 793022 ( INDIA )}

\date{\today}

\begin{abstract}
 In this paper, we investigate the validity of the generalized
second law of thermodynamics (GSLT) in  the DGP brane world when
universe is filled with interacting two fluid system: one in the
form of cold dark matter  and other is holographic dark energy.
The boundary of the universe is assumed to be enclosed by the
dynamical apparent horizon or the event horizon. The universe is
chosen to be homogeneous and isotropic  FRW model and the validity
of the first law has been assumed here.

\end{abstract}

\pacs{98.80.Cq, 98.80.-k}

\maketitle

\textbf{Keywords }: DGP brane-world; holographic dark energy;
generalised second law of thermodynamics (GSLT).

\section{\normalsize\bf{Introduction}}

Astrophysical observations made at the turn of the last century
[1] show conclusive evidence for acceleration in the late
universe, which is still a challenge for cosmologists. It shows
beginning of accelerated expansion in the recent past. It is found
that cosmic acceleration is driven by some invisible fluid having
its gravitational effect in the very late universe. This unknown
fluid has distinguishing feature of violating strong energy
condition(SEC) being called dark energy (DE)[2]. Various models
have been proposed to solve this problem. A comprehensive review
of these models is available in [3].

In the race to investigate a viable cosmological model, satisfying
observational constraints and explaining present cosmic
acceleration, brane-gravity was introduced  and brane-cosmology
was developed. A review on brane-gravity and its various
applications with special attention to cosmology  is available
in[4].

 A simple and well studied model of
brane-gravity (BG) is the Dvali-Gabadadze-Porrati(DGP) braneworld
model[5]. In this model our 4-dimensional world is a FRW brane
embedded in a 5-dimensional Minkowski bulk. It explains the origin
of DE as the gravity on the brane leaking to the bulk at large
scale.
 On the
4-dimensional brane the action of gravity is proportional to
$M_{P}^{2}$ whereas in the bulk it is proportional to the
corresponding quantity in 5-dimensions. The model is then
characterized by
 a cross over length scale
     $$ r_{c}=\frac{M_{P}^{2}}{2M_{5}^{2}}$$
such that gravity is 4-dimensional theory at scales $ a\ll r_{c}$
where matter behaves as pressureless dust but gravity
\textit{leaks out} into the bulk at scales $ a\gg r_{c} $ and
matter approaches the behaviour of a cosmological constant.

    In this conceptual set up, one of the important questions
concerns the thermodynamical behaviour of an accelerated expanding
universe driven by DE. The first hint on the connection between
general relativity and thermodynamics was given by Bekenstein in
1973[6]. He outlined the laws of thermodynamics in the presence of
black holes which turned out to be equivalent to the laws of black
hole mechanics [7]. Study of gravitational thermodynamics in an
accelerating universe has been a strong candidate and has been
addressed to in many papers
 based on General relativity (GR)[8]. The reason of interest in this
  subject is two fold (i)it
is natural to study thermodynamical aspect of accelerating
universe and (ii) the astonishing result for phantom obtained in
[9] which has either negative temperature or negative entropy.
This is another
 problem of phantom cosmology like big-rip
  singularity for which some viable solution are proposed in  [10]. The main problem
  of studying
thermodynamics of the Universe  is to define
 the entropy and temperature on the boundary of the universe. Generally the entropy and hence
 the temperature is taken from black hole physics but in other gravitational theories(such as $f(R)$ gravity)
 some correction terms may be needed.

 Motivated by the profound connection
between black hole physics and thermodynamics, in recent times
there has been some deep thinking on the relation between gravity
and thermodynamics. A pioneer work in this respect was done by
Jacobson who disclosed that Einstein's
 gravitational field equation
  can be derived from the relation between horizon area and entropy together with
Clausius relation $\delta Q = T\delta S $ [11]. Some recent
discussion on the connection between gravity and thermodynamics on
various gravity theories can be found on [12]. Recently this
connection between gravity and thermodynamics has been extended to
brane world scenarios [13].~In ref [14] it is shown that apparent
horizon entropy extracted through connection between gravity and
first law of thermodynamics satisfies the generalised second law
of thermodynamics (GSLT) in DGP warped brane. In General
Relativity (GR) frame work the authors of [15] have shown in
contrast to the case of the apparent horizon,~both first and
second law of thermodynamics breakdown if one considers boundary
of the universe to be the event horizon. Essentially the two
horizons have significant difference both from geometrical and
physical point of view. The cosmological event horizon exists in
accelerating universe while it may not exist in the standard big
bang model. From the thermodynamical point of view the universe
bounded by the apparent horizon is a Bekenstien system having well
defined entropy and temperature while universe bounded by the
event horizon may not be a Bekenstein system and hence temperature
and entropy are not well defined.

The other way to approach to the problem of DE arises from
holographic principle which states that the number of degrees of
freedom for a system within a finite region should be finite and
is bounded by the area of its boundary. As in ref [16] one obtains
holographic energy density  as
$$ \rho_{D} = 3 c^{2}M_{P}^{2}L^{-2}$$
 where $L$ is an IR cut-off in units $M_{P}^{2}=1$.
Li shows that [17] if we choose L as the radius of the event
horizon we can get the correct equation of state and get the
desired accelerating universe.

It may be noted that in literature, standard DGP model has been
generalized to (i) LDGP model by adding a cosmological constant
[18],(ii) QDGP model by adding quiessence perfect fluid [19],
(iii) CDGP by Chaplygin gas [20](iv) SDGP by a scalar field [21].
In [22] the DGP model  has been analysed by adding Holographic
dark energy (HDE).

In a recent paper[23], validity of GSLT has been studied on
 event horizon for interacting  DE. Assuming first law of
 thermodynamics on the event horizon, they have found conditions
 for validity of GSLT in both cases when FRW universe is filled with interacting
 two fluid system- one in the form of cold dark
matter and the other is either holographic dark energy or new
agegraphic dark energy.  In  [24], we have investigated the
validity of GSLT of universe in the  DGP brane world. The boundary
of the universe was assumed to be enclosed by the dynamical
apparent horizon or the event horizon. In the present paper, we
extend this investigation to interacting holographic dark energy
model in DGP brane world. The matter in the universe is taken in
the form of interacting two fluid system- one component is cold
dark matter(CDM) and the other is in the form of HDE.

The paper is organized as follows : Section 2 deals with
interacting HDE in the DGP brane model while validity of GSLT has
been examined for apparent  and event horizon and conclusion are
written pointwise in section 3.

\section{\normalsize\bf{Interacting Holographic dark energy in DGP model:}}

In  flat, homogeneous and isotropic brane the Friedmann equation
[5]  in DGP model is given by

$$ H^{2}=\Big(\sqrt{\frac{\rho}{3}+\frac{1}{4r_{c}^{2}}}+\epsilon
\frac{1}{2r_{c}}\Big)^{2}       \eqno(2.1)$$

 or equivalently

$$ H^{2}-\epsilon \frac{H}{r_{c}}=\frac{\rho}{3}  \eqno(2.2)$$

 where $ H=\frac{\dot a}{a}$ is the Hubble parameter,~$\rho$ is the
 total cosmic fluid energy density and
  $ r_{c}=\frac{M_{P}^{2}}{2M_{5}^{2}}$ is the
crossover scale which determines the transition from 4D to 5D
behavior and $\epsilon = \pm 1 $.\\(For simplicity we are using $8
\pi G =1$ )

For $ \epsilon = 1 $, we have standard DGP(+) model which is self
accelerating model without any form of dark energy, and effective
$\omega$ is always non phantom. However for $ \epsilon = - 1 $, we
have DGP(-) model which does not self accelerate but requires dark
energy on the brane. It experiences 5D gravitational modifications
to its dynamics which effectively screen dark energy.

Here we take $\rho = \rho_{m}+ \rho_{D}$ where $\rho_{m}$ is the
energy density of cold dark matter(CDM) and $ \rho_{D} $ is the
energy density of HDE.

The Friedmann eq.(2.2) can be written in following effective
Einstein form

 $$ H^{2}=
\frac{1}{3}(\rho_{m}+\rho_{eff})     \eqno(2.3)$$

$$ \dot H =-\frac{1}{2}\left\{\rho_m +(\rho_{eff} +p_{eff})\right\}               \eqno(2.4)$$
 where
$\rho_{eff}$ is the effective energy density given by

$$\rho_{eff}= \rho_{D} +\epsilon \frac{3H}{r_{c}}   \eqno(2.5)$$

and $ p_{eff}$ is the effective pressure given by

$$ p_{eff}=  p_{D}-\epsilon\frac{3H}{r_{c}}-\epsilon\frac{\dot H}{r_{c}H}   \eqno(2.6)$$

The individual   conservation  equation for effective DE and CDM
are respectively given by

$$\dot\rho_{eff}+3H(1+\omega_{eff})\rho_{eff}=-Q    \eqno(2.7)$$
and
$$      \dot \rho_{m}+3H\rho_{m} = Q  \eqno(2.8)$$

where $Q=\Gamma \rho_{D}$ is called interaction term [23] and the
decay rate $\Gamma$ corresponds to conversion of dark energy to
dust (CDM). Following [23], if we define

$$ \omega_{eff}^{(i)}=
\omega_{eff} +\frac{\Gamma}{3H}+ \frac{\dot H}{r_c H}
~~~~~~~~\textrm{and}~~~~~~~~ \omega_{m}^{(i)}=-\frac{\Gamma u}{3H}
                                                         \eqno(2.9)$$

then the above conservation equations can be written in
non-interacting form as

$$\dot\rho_{eff}+3H(1+\omega_{eff}^{(i)})\rho_{eff}= 0    \eqno(2.10)$$
and
$$      \dot \rho_{m}+3H(1+\omega_{m}^{(i)})\rho_{m} =0   \eqno(2.11)$$

where $u=\frac{\rho_D}{\rho_m}$ is the ratio of energy densities.

Also using eq.(2.5) in eq. (2.10) the actual energy conservation
for DE is

$$\dot \rho_{D}+3H(1+\omega_{D}^{(i)})\rho_{D}=0 \eqno(2.12)$$

where

$$\omega_{D}^{(i)}=\omega_{D}+ \frac{\Gamma}{3H}     \eqno(2.13)$$

Combining eqs.(2.10) and (2.11), we get

$$\dot\rho_{t}+3H(\rho_{t}+ p_t)= 0    \eqno(2.14)$$

where   $$\rho _t= \rho_m + \rho_{eff}~~~~  \textrm{and}~~~~~
p_t=p_m+p_{eff}=\omega_{m}^{(i)} \rho_m +
\omega_{eff}^{(i)}\rho_{eff}  \eqno(2.15)$$

\section{\normalsize\bf{The Generalized Second Law of thermodynamics:}}

In this section we examine the validity of GSLT on  3-DGP brane.
Let us consider a region of FRW universe
   enveloped by the  horizon and assume that the region bounded by the horizon act as a thermal
   system with boundary defined by the  horizon and is filled with a perfect fluid of energy density
    $\rho_{t}$ and pressure $p_{t}$ given by eq. (2.15)

Gravity on the brane does not obey Einstein theory, therefore
usual area formula for the black hole entropy may not hold on the
brane. So we extract the entropy of the event horizon by assuming
the first law of thermodynamics on event horizon [23].

The amount of energy crossing the  horizon in time $dt$ has the
expression

$$ -dE = 4\pi R_{h}^{3}H(\rho_{t}+p_{t})dt        \eqno(3.1)  $$

where $R_{h}$ is the radius of the  horizon.

 So from the first law of thermodynamics we have

$$ \frac{dS_{h}}{dt}= \frac{4\pi R_{h}^{3}H}{T_{h}}\Big[(1+\omega_{m}^{(i)})
\rho_{m}+(1+\omega_{eff}^{(i)})\rho_{eff}\Big]  \eqno(3.2) $$

where $S_{h}$ and $ T_{h}$ are the entropy and temperature of the
 horizon respectively.

  Using Gibb's equation [8],

 $$ T_{h}dS_{I} = dE_{I} + p_{t}dV $$

 we obtain the variation of the entropy of the fluid  inside  the
horizon as

$$T_h dS_{I} = d(\rho_m+\rho_D)V+ (\rho_m+\rho_D + p_{D})dV    \eqno(3.3)$$ where $S_{I}$ and
$E_{I}$ are the entropy and energy of the matter distribution
inside the  horizon. Here we assume as in ref [25] the temperature
of the source inside the  horizon is in equilibrium with the
temperature associated with the horizon.

So starting with $ E_{I}=\frac{4}{3}\pi R_{h}^{3}\rho_{t}$ and $ V
= \frac{4}{3} \pi R_{h}^{3}$ and using eqs.(2.11),(2.12b) and
(3.3) and after some simplification one gets

$$\frac{dS_{I}}{dt}=  -\frac{4\pi R_{h}^{3}H}{T_{E}}\Big[\{\rho_m (1+\omega_{m}^{(i)})+\rho_{D}
(1+\omega_{D}^{(i)})\rho_{eff}\Big]+\frac{4\pi
R_{E}^{2}}{T_E}\frac{dR_E}{dt}\Big[\rho_m+\rho_D(1+\omega_D)\Big]
   \eqno(3.4) $$

Adding eqs. (3.2) and (3.4), one gets the resulting change of
entropy

$$ \frac{dS_{h}}{dt} + \frac{dS_{I}}{dt}= \frac{4\pi R_{h}^{2}}{T_{h}}\Big[-\frac{\epsilon
\dot H R_{h}}{r_c}+ \{\rho_m+\rho_D(1+\omega_{D})\}\Big]
\eqno(3.5)$$

We shall now examine the validity of GSLT i.e.,

$$\frac{dS_{h}}{dt} + \frac{dS_{I}}{dt}\geq  0$$

for  apparent and event horizon respectively.

\subsection{\normalsize\bf{Universe bounded by apparent horizon}}
Here $$R_h=R_A=\frac{1}{H}   \eqno(3.6)$$

So $$\frac{dR_A}{dt}=-\frac{\dot
H}{H^{2}}=\frac{1}{2H^2}\left\{\rho_m +(\rho_{eff}
+p_{eff})\right\}         \eqno(3.7)$$

Hence eq. (3.5) simplifies to
$$ \frac{dS_{tot}}{dt}= \frac{4\pi R_{A}^{2}}{T_{A}}\Big[\epsilon\frac{
 R_{A}}{2 r_c}+
 \frac{1}{2H^2}\{\rho_m+\rho_D(1+\omega_{D})\}\Big]\Big\{\rho_m +\rho_{eff}(1+\omega_{eff})
  \Big\} \eqno(3.8)$$

\subsection{\normalsize\bf{Universe bounded by event horizon}}
In this case
$$R_h=R_E$$
Here due to holographic nature of the DE the energy density of the
holographic matter can be written as

$$\rho_{D} = \frac{3 c^{2}}{R_{E}^{2}}      \eqno(3.9) $$

So using the conservation equation (2.12) the time variation of
the event horizon can be written as [26]
$$\dot R_E= \frac{3}{2}H R_E (1+\omega_{D}^{(i)})  \eqno(3.10)$$

Hence eq. (3.5) now becomes
$$ \frac{dS_{E}}{dt} + \frac{dS_{I}}{dt}= \frac{4\pi R_{E}^{2}}{T_{E}}\Big[\frac{\epsilon
 R_{E}}{2 r_c}\Big\{\rho_m +\rho_{eff}(1+\omega_{eff})
  \Big\}+ \frac{3}{2}HR_E(1+\omega_{D}+\frac{\Gamma}{3H})
  \{\rho_m+\rho_D(1+\omega_{D})\}\Big] \eqno(3.11)$$

  The following conclusion we can draw from the variation of the
  total entropy given by equations (3.8) or (3.11):\\

  {\bf (a)} The entropy variation does not depend on the
  interaction in the case of apparent horizon while in the case of
  event horizon there is dependence on interaction through $ \dot
  R_E$ due to the choice of the energy density for holographic
  dark energy given by eq. (3.9)

  {\bf (b)} If we put $r_c\rightarrow\infty$(this happens  at
  early epoch) i.e., neglect the brane effect then GSLT will always
  be satisfied for apparent horizon while GSLT will not be trivial
  for event horizon as it depends on the interaction term
  $\Gamma$.

  {\bf (c)} In deriving GSLT we do not need  any specific choice
  for entropy and temperature at the horizon.

  For future work it will be interesting to determine the form of
  entropy and temperature at the horizon in brane scenario.

  {\bf Acknowledgement:}\\
The work is done during a visit to IUCAA, Pune, India. The authors
are thankful to IUCAA for warm hospitality and facility of doing
research works. One of the author (S.C) is thankful to CSIR for a project on brane
cosmology.\\\\

{\bf References:}\\
\\
$[1]$ Perlmutter,~ S. ~J.  $et$ $al.$(1999), Astrophys. J. {\bf
517},565 (1998) astro-ph/9812133;\\Spergel,~D. ~N.  $et$ $al$,
Astrophys J. Suppl. {\bf 148} (2003)175[ astro-ph/0302209] and references therein.\\\\
$[2]$Riess,~ A. ~G. $et$ $al$,(2004), Astrophys. J. {\bf 607}, 665
[ astro-ph/0402512].\\\\
$[3]$Copeland,~E.~J., Sami,~M. and  Tsujikawa,~S.  Int.
J.Mod.Phys.D,
{\bf 15},(2006)1753 [hep-th/0603057] and references therein. \\\\
$[4]$Rubakov, ~V. ~A.,(2001), Phys. Usp. {\bf 44}, 871
[hep-ph/0104152];\\~Maartens,~ R.,(2004) Living Rev. Relativity,
{\bf 7}, 7,  [(2003) gr-qc/0312059];\\ ~Brax,~ P. et al,(2004),
Rep. Prog.Phys. {\bf 67}, 2183 [hep-th/0404011]; \\~Csa$^{\prime}$ki, C.,(2004) [hep-ph/0404096].\\\\
$[5]$ Dvali,~G.~R. , Gabadadze,~G. and Porrati, ~M. (2000),
Phys.Lett. {\bf B 485}  208 [hep-th/000506];\\~ Deffayet,~D.
(2001), Phys.Lett. {\bf B 502}  199;\\~ Deffayet,~D., Dvali,G.R.
and
Gabadadze,~G.(2002), Phys.Rev.{\bf D 65}  044023 [astro-ph/0105068].\\\\
$[6]$Bekenstein,~ J.~ D (1973) Phys. Rev. \textbf{D 7} 2333\\\\
$[7]$ Hawking,~ S.~ W. 1975 Commun. Math. Phys. \textbf{43
}199.\\\\
$[8]$ ~Davies,~P.~C.~W.,(1987)~Class.~Quan.~Grav {\bf
4},L225;~Class.~Quan.~Grav {\bf  5},1349(1988);\\Izquierdo,~ G.
and Pavon,~ D.   ,(2006), Phys. Lett. {\bf B 633},  420;\\~ Wang,
~B.,~ Gong, Y. and  Abdalla, E.  (2006), Phys.Rev.{\bf D 74}
083520
 [gr-qc/0511051];\\~Akbar,M. and Cai, R.G.   (2006), Phys.Rev.{\bf D 73}  063525
[gr-qc/0512140];\\~ Sadjadi, M. H.   (2007), Phys.Rev.{\bf D 75}
084003 [hep-th/0609128].\\\\
$[9]$ Brevik,  I.,~Nojiri, S.,~Odinstov, ~S.~D. and
~Vanzo,~L.~(2004) Phys.Rev.{\bf D 70}~043520,\\\\
$[10]$ Srivastava, S.K.; (2005), Phys. Lett. B {\bf 619} 1
 [astro-ph/0407048].\\\\
$[11]$ Jacobson,~ T. (1995), Phys.Rev.Lett. {\bf 75} 1260
[gr-qc/9504004].\\\\
$[12]$  Eling, ~C., Guedens,~R. and Jacobson,~ T. (2006),
Phys.Rev.Lett. {\bf 96} 121301;\\~ Akbar,~M. and Cai, ~R.~G.
(2006),\\\\
$[13]$ Cai, R.G. and Cao, L. M.(2007) , Nucl. Phys. {\bf B 785},
  135;\\~ Sheykhi, A. , Wang, B. and Cai, R.G. (2007) , Nucl. Phys. {\bf B 779},
  1;\\~ Sheykhi, A. , Wang, B. and Cai, R.G.  (2007), Phys.Rev.{\bf D 76}  023515 [hep-th
  /0701261];\\ ~Sheykhi, A. (2009)
  JCAP {\bf 0905}, 019 (2009).  \\\\
$[14]$ Sheykhi, A. , Wang, B. [arXiv:0811.4478]\\\\
$[15]$ ~ Wang, B., Gong, Y. and  Abdalla, E.  (2006),
Phys.Rev.{\bf D 74}  083520 [gr-qc/0511051].  \\\\
$[16]$ Cohen, A. G., Kaplan, D. B., and  Nelson, A. E. (1999),
Phys.Rev.Lett.{\bf 82} 4971.\\\\
$[17]$ Li, M. (2004), Phys. Lett. {\bf B  603},  01.\\\\
$[18]$  ~Lue,A. and ~Starkman,G.~D.(2004)
  Phys.\ Rev.\  D {\bf 70}, 101501
  [arXiv:astro-ph/0408246].   \\\\
$[19]$ ~Chimento,L.~P., ~Lazkoz, R., ~Maartens, R. and ~Quiros,
I.(2006)
  JCAP {\bf 0609}, 004  [arXiv:astro-ph/0605450].\\\\
$[20]$ ~Bouhmadi-Lopez, M. and ~Lazkoz, R.(2007)
  Phys.\ Lett.\  B {\bf 654}, 51
  [arXiv:0706.3896 (astro-ph)].\\\\
$[21]$~Zhang,~H. and Zhu,~Z.~H.(2007) Phys.Rev.{\bf D
75},023510.\\\\
$[22]$ Wu,~X., Cai,~R.~G. and Zhu,~Z.~H (2008)Phys.Rev.{\bf D
77},043502.
\\\\
$[23]$ Mazumder, N. and Chakraborty, S.(2010),
  [arXiv:1005.5589 [gr-qc]].
\\\\
$[24]$ Dutta,~J. and ~Chakraborty, S., Gen.\ Rel.\ Grav.\
$doi:10.1007/s10714-010-0957-9$
\href{http://www.slac.stanford.edu/spires/find/hep/www?irn=8613044}{SPIRES
entry}\\\\
$[25]$ Saridakis,F.N.,Gonzalez-Dyaz,P.F.and Siguenza,
C.I.[arXiv:0901.1213/astro-ph];\\~Saridakis,F.N.,Gonzalez-Dyaz,P.F.and
Siguenza, C.I. Nucl. Phys.(2004) {\bf B 697}, 363;\\~Pereira,
S.H.and Lima,J.A.S. (2008), Phys. Lett. {\bf B  669}, 266.\\\\
$[26]$ ~Mazumder, N., and ~Chakraborty, S.
  Gen.\ Rel.\ Grav.\  {\bf 42}, 813 (2010)
  [arXiv:1005.3403 [gr-qc]].

\end{document}